# Flow and filtration imaging of single use sterile membrane filters

M. Wiese[a], C. Malkomes[a], B. Krause[c], M. Wessling[a,b,*]

[a]*Chemical Process Engineering, RWTH Aachen University, Forckenbeckstrasse 51, D-52056 Aachen, Germany*
[b]*DWI-Leibniz-Institute for Interactive Materials, Forckenbeckstrasse 50, D-52056 Aachen, Germany*
[c]*Baxter Membranes & Devices R&D, Holger-Crafoord-Strasse 26, D-72379 Hechingen, Germany*

**Abstract**

Sterile filters incorporating endotoxin adsorption function emerged recently to pretreat dialysate liquids fed to a hemodialysis filtration process. Their application significantly enhances the survival rate during dialysis treatment as they function as steril filters as well as an endotoxin adsorber. Little is known about the fluid flow distribution in such single use membrane modules. We report a detailed analysis of the local 3D flow field distribution in such membrane modules using magnetic resonance flow imaging. Next to pure water filtration representing the application case of endotoxin adsorption from an already pure liquid, we also used the module as a filtration device rejecting for instance colloidal silica. Such experiments performed in-situ allow the quantification of cake layer development and its concomitant redistribution of the flow field. Particularly novel is the quantification of the time evolution of local permeate flux distribution. These detailed insights of this study encourage the use of flow-MRI when designing and applying new membrane module configurations.

## 1. Introduction

The prevalence of hollow fiber based hemodialysis reached 2067 per million population in 2014 [1] with an increasing tendency. This trend underlines the issue of a global public health problem [2, 3] and the necessity for continuous improvements in medical devices enabling the lifesaving treatment. Recently, single-use microfiltration membrane based devices act as sterile filters as well as endotoxin adsorbers for dialysate pre-treatment. They (a) eliminate any bacterial contamination and (b) reduce the level of endotoxins in the dialysate to a minimum and thus significantly increase the patients survivability [4]. Considering the enormous number of these filtration devices used annually, it becomes evident that a comprehensive understanding of its function is inevitable. To utilize the membrane filtration and adsorption capacity effectively, the flow distribution over the membrane should be equal over the whole membrane and should not alter in time.

Non-invasive observation techniques such as X-ray computed tomography (CT) and nuclear magnetic resonance (NMR) are proven techniques for the insightful analysis of structure and flow in membrane devices [5]. Specifically in the field of medical membrane devices, non-invasive observation has already been applied to hemodialyzers to track fluid concentrations within the fibers and to image the packing density [6]. Observed by X-ray CT, the wet-out procedure in hemodialyzers revealed that water preferably flows around areas with high fiber packing and by thus traps air in regions with low packing density [7]. For the sterile filter and endotoxin adsorber module little is known so far about the flow distribution and the filtration performance.

This paper presents a comprehensive in-situ flow monitoring methodology based on NMR to determine (a) the macroscopic flow distribution in the feed channel as well as (b) the local flow distribution in permeate channels. To showcase the strength of local permeate flow distribution we expand the analysis towards the quantification of the transient 3D-resolved flow distribution during frontal colloidal filtration.

## 2. Background on MRI

Magnetic resonance imaging (MRI) is a suitable tool to access the membrane module's inner details [8]. Magnetic resonance imaging is a non-invasive technique particularly able to analyze opaque and non-transparent modules. We have been using flow-MRI to analyze hydrodynamic conditions in 3D printed modules. The latter becomes highly important since 3D printing enables freedom of design in module fabrication that leads to complex devices with unprecedented geometry [9]. Staggered herring bones are such complex geometries which have been successfully embedded in microfluidic devices by 3D printing to introduce vorticity and have been analyzed in-situ and online by 3D flow-MRI [10].

One of the first studies to successfully combine MRI and membrane separation dealt with the quantitative flow analysis in axial direction in a single and multiple fiber bioreactor module [11]. Imaging of deposited silica particles in single tubular membrane module showed the time-dependent thickness of the concentration polarization layer severely influenced by gravity [12]. Various module designs of hollow fiber bundles used in membrane distillation were analyzed by low-field MRI revealing the limits of low-field MRI with respect to low signal-to-noise ratios (SNR) but also its capabilities when it comes to flow analysis in the module [13]. In tubular membrane systems, MRI showed to be an effective tool to visualize the deposition of silica particles and the subsequent decrease in flow in the lumen of the fiber [14]. In terms of module performance, first correlations were made to access the local cumulative flux as a function of the vertical position [14]. In a study containing two different silica suspensions MRI visualized the cake layer formation on the fiber in dependency of the fiber shape while highlighting the different deposition behavior [15]. Flow-MRI investigation in the field of flat sheet membrane modules primarily targeted on the evolution of biofouling in NF membrane modules [16] regarding the feed spacer impact [17] and various cleaning steps to remove the biofilm [18].

In contrast to previous studies, our flow-MRI investigation resolves the spatial flow field of the sterile membrane filter using plane thicknesses of only several millimeters. The latter allowed us to focus on local flow field changes including the local flux. Based on these flow-MRI data, we can quantitatively derive the feed and shell volume flows and the active membrane area.

## 3. Materials & Methods

### 3.1. Dead-end filtration module

The membrane module investigated in our studies is a flatsheet membrane module developed by Gambro Dialysatoren GmbH. Used as a pre-filter it performs a gatekeeper function and retains any endotoxins before contacting the dialysate stream with the sterile hemodialysis module. The inserted membrane is positively charged and hydrophilic with a mean pore size diameter of 0.2 $\mu$m. Figure 1 shows a picture of the membrane filter and its module composition (lid, membrane and bottom part). For a higher mechanical stability of the membrane filter the lid features three splines aligned in the z-direction that influence the fluid dynamics within the module. Further prominent module design features comprise the sharp incline in the inlet and the steadily growing height of the permeate channels against the z-axis (Figure 1b). The rectangular shaped permeate channels that hinder the membrane from bending and the structure of the splines are shown in Figure 1c. The membrane is welded right on top of the permeate channels to avoid transversal short cut flows.

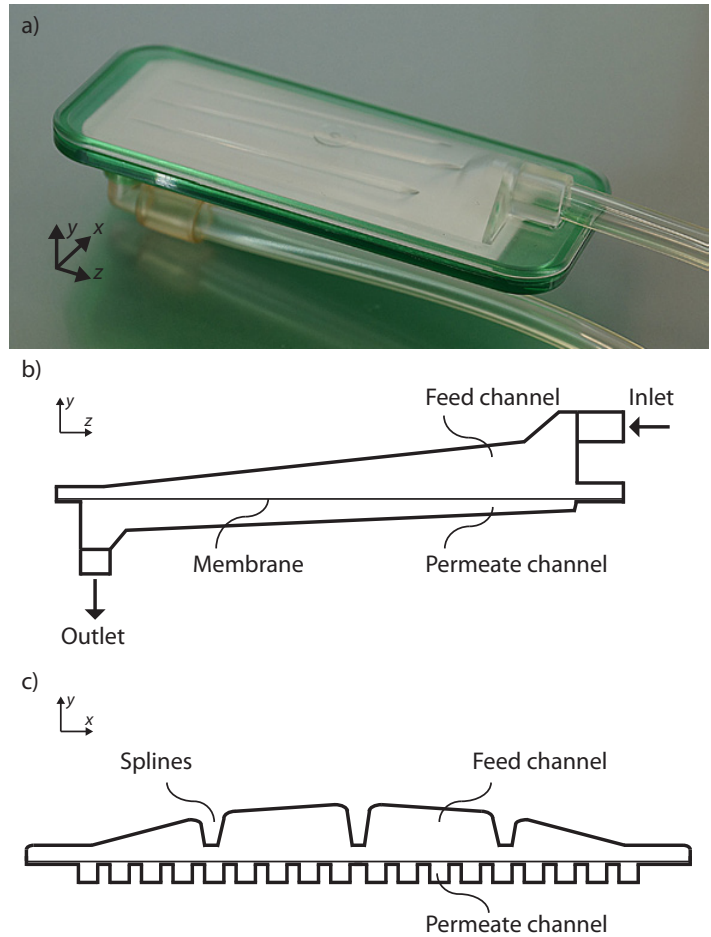

Figure 1: (a) Picture of Baxter's sterile membrane filter with sketches of the (b) *zy*-view and (c) *xy*-view. Sketches serve the purpose to highlight the design features of the module and for clarification of the NMR images.

*3.2. Model suspension*

For the cake layer formation studies, aqueous dispersions of amorphous colloids of spherical shape were used. The silica particles SYLOID C803 ($SiO_2$) have a particle size of 3.4 $\mu m$ to 4.0 $\mu m$ and a density of 2.2 g/ml at 20°C [19]. For filtration experiments a concentration of 4.3 g/l was used which is equal to approximately 0.2 vol-% [14].

*3.3. Experimental setup and filtration procedure*

Pure water and silica filtration experiments were performed with the setup shown in Figure 2. At the core of the setup is the NMR tomography system in which the membrane filter was aligned parallel to the main magnetic field direction (z-axis, see Figure 1). The buffer solution consisted of de-ionized water with 1 g/l copper sulfate and was kept constant at 29°C to avoid interference with the magnetic field of tomography system. To be close to the application range of the membrane filter a volume flow of 21 *ml*/*min* was adjusted and fed to the module via two continuously and pulsation free working syringe pumps. Right before the membrane module a pressure sensor was installed to track the pressure in the membrane module and to determine the transmembrane pressure. For the silica filtration experiments an additional syringe pump mixed the buffer solution with the silica suspension of equal copper sulfate concentration (dashed line, Figure 2). Since the pore size of the membrane is smaller than the average particle size of the silica particles total separation is guaranteed. For every series of filtration experiments we used one single membrane module which has been discarded afterwards.

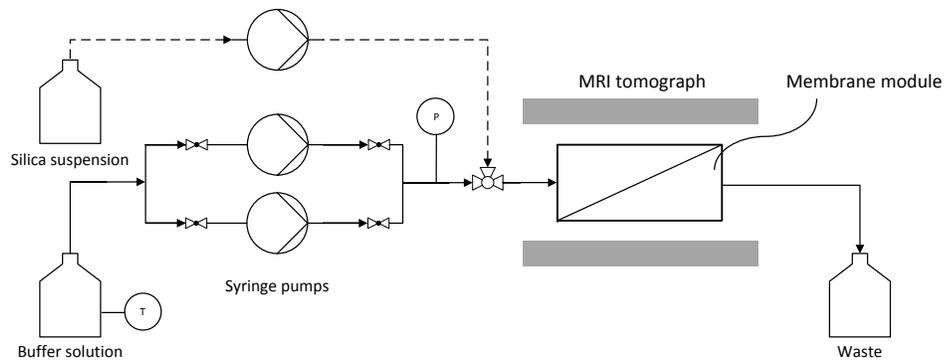

Figure 2: Filtration process with combined online NMR tomography.

In-situ NMR imaging of the silica filtration comprised alternating steps of filtration with and without applied imaging gradients (Figure 3). While running the setup with the buffer solution ($t_0$) we started adding the silica suspension to the module for 5 *min* of filtration. The spin echo pulse sequence took about 4 *min* to take a first MR image without velocity information. In order to derive qualitative flow-MRI images we immediately took a second image with an applied magnetic velocity gradient. Consequently, we gathered two consecutive MRI images showing the cake layer evolution (Figure 9) and one corresponding velocimetry image (Figure 10). The combined filtration and imaging cycle was repeated for approximately one hour.

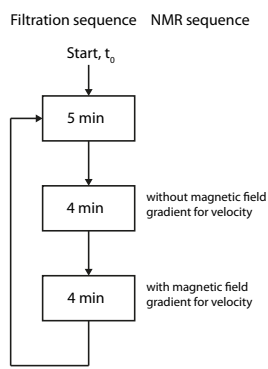

Figure 3: Silica filtration procedure adapted to the NMR measurement conditions.

*3.4. NMR imaging*

NMR imaging is a non-invasive technique that exploits the nuclear magnetic properties of the protons. An object brought into a NMR device will cause the magnetization of the contained protons to align along the static magnetic field. Implemented radio frequency coils establish an oscillating external magnetic field in the radiofrequency regime of the protons to tip them out of their equilibrium state. As soon as the external magnetic field is shut off the protons rotate into their equilibrium emitting a signal. Subsequent Fourier transformation translates this signal to create images. With the help of gradient coils a linearly varying magnetic field is applied which can be directly referred to the local position and velocity (Figure S1). All NMR measurements presented in this study were performed on a Magritek low-field NMR tomography system with a field strength of 0.56 T and a birdcage size of 60mm that operates at a Larmor frequency of 23.8 MHz. Depending on the object to image the NMR specific parameters highly vary. We used the spin-echo pulse sequence for imaging, flow visualization and cake layer observation [20]. The according NMR parameters are summarized in Table 1. Signal to noise ratios were at all times higher than 10.

**Table 1**

NMR parameters for flow visualization and cake layer observation

| NMR settings | Flow visualization | Cake layer observation |
| --- | --- | --- |
| Sequence type | SE-2D-vel | SE-2D-vel |
| Data acquisition [mm:ss] | 51:12 | 4:16 |
| Repetition time $T_R$ [ms] | 600 | 500 |
| Echo time $T_{echo}$ [ms] | 35 | 35 |
| Number of scans $N$ | 20 | 4 |
| Fielf of View ($x$ x $y$ x $z$) [mm x mm x mm] | 50 x 32 x 100 | 50 x 32 x 100 |

*3.5. Data processing with Matlab*

Post-processing was realized by importing the measured MRI data from Magritek's software PROSPA to Matlab. In Prospa the MRI information are saved pixel by pixel in matrices. Using Matlab we derived corresponding contour plots, vector plots and the procedure for the mass balance which we applied for local flux validation. The contour and vector plots in the results section were derived from predefined Matlab commands. Additionally, data smoothing helped to capture the distinctive patterns in the MRI data while clearing out local outliers coming from noise or MRI artifacts (6). For the volume flow analysis in Chapter 4.3 the following procedure was adapted.

*3.6. Volume flow calculation*

We established a Matlab method to quantitatively determine the volume flow within a predefined area assuming a constant density of the fluid throughout the measurements. Based on *z*-velocity MRI data this method was then applied to support the local flux measurements. In that case, we compared the fluid permeating through the membrane with the difference of volume flow entering and leaving the volume cell (Figure 4a):

$$\dot{V}_{permeate} = \dot{V}_{out} - \dot{V}_{in} \qquad (1)$$

The volume flows $\dot{V}_{out}$ and $\dot{V}_{in}$ were derived from MRI measurements following

$$\dot{V}_{in,out} = \bar{v}_{in,out} \cdot A_{channel} \qquad (2)$$

with the corresponding median velocity $\bar{v}_{in,out}$ and cross sectional area of the permeate channel $A_{channel}$. Only pixels containing velocity information were considered to be relevant and had the size of

$$A_{Pixel} = \frac{FoV_{Read}}{N_{Pixel,read}} \cdot \frac{FoV_{Phase}}{N_{Pixel,phase}} = \Delta y \cdot \Delta x \qquad (3)$$

The sum of all relevant pixels then led to the total area of the permeate channel $A_{channel}$. In Figure 4b one pixel is exemplary illustrated. The median velocity was derived by row and column multiplication of the imported $z$-velocity matrix

$$\begin{pmatrix} 1 & \cdots & 1 \end{pmatrix} \cdot \begin{pmatrix} v_{11} & v_{12} & \cdots & v_{1j} \\ v_{21} & v_{22} & \cdots & v_{2j} \\ \vdots & \vdots & \vdots & \vdots \\ v_{i1} & v_{i2} & \cdots & v_{ij} \end{pmatrix} \cdot \begin{pmatrix} 1 \\ \vdots \\ 1 \end{pmatrix} = v_{total} \qquad (4)$$

and

$$\bar{v} = \frac{v_{total}}{N_{Pixel}} \qquad (5)$$

Please note that the amount of pixels doubled due to zero filling and that all pixels have the same constant size.

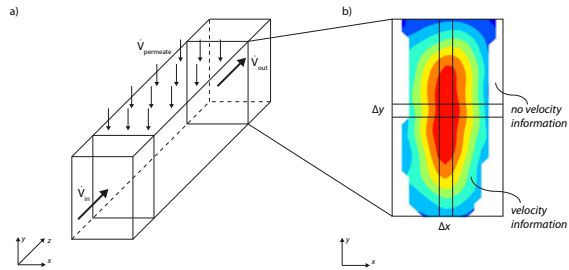

Figure 4: Scheme of the volume flow balancing within the permeate channels.

## 3.7. Drag and gravitational force

During filtration particles in a colloidal suspension experience several forces that ultimately determine their direction of movement. In the simplest case we can reduce the number of forces to the drag force $F_D$ and the gravitational force $F_G$. The drag force $F_D$ also known as Stokes' drag is calculated by

$$F_D = 6 \cdot \pi \cdot \eta \cdot R \cdot v \tag{6}$$

where $\eta$ is the dynamic viscosity, $R$ the radius of the sphere and $v$ as the velocity the sphere experiences. The latter is determined as an average velocity by

$$v = \frac{\dot{V}}{A_{membrane}}. \tag{7}$$

The gravitational force is described by

$$F_G = \frac{4}{3} \cdot \pi \cdot R^3 \cdot (\rho_{particle} - \rho_{liquid}) \cdot g \tag{8}$$

with $g$ as the gravitational acceleration and $\rho$ as the density of the particle and liquid respectively. Balancing both forces leads to the main direction of motion of the particle. In this study we exclude other driving forces due to the fact that (a) we did not apply any external fields, (b) the silica particles are not influenced by the magnetic field of the tomography and (c) salt effects are neglected since we use de-ionized water. Alternatively, the influence of the fluid flow can be evaluated using the Stokes number

$$Stk = \frac{t_0 \cdot u_0}{d_p} \tag{9}$$

with $d_p$ as the particle diameter, $u_0$ as the fluid velocity and $t_0$ as the relaxation time of the particle. The latter is defined as

$$t_0 = \frac{\rho_p \cdot d_p^2}{18 \cdot \eta_l} \tag{10}$$

with $\rho_p$ as the particle's density and $\eta_l$ as the viscosity of the liquid. If the $Stk$ number is much smaller than 1, the particle follows the streamlines of the fluid and inertial forces are negligible. For $Stk \gg 1$ inertia dominates the particle's movement. This simple analysis is valid for low $Re_p \ll 1$. As described later, we find this analysis to be justified as $Re_p$ becomes

$$Re_p = \frac{\rho_l \cdot d_p \cdot v}{\eta_l} = 7.4 \cdot 10^{-4} \ll 1. \tag{11}$$

## 4. Results & Discussion

MRI measurements visualized the interior structure of the membrane filter highlighting the membrane, the inlet and the module's lid. With respect to fluid dynamic analysis, flow-MRI completed our data set to show the effect of flow distribution on the filtration behavior. Points of interest are thereby the flow evolution along the module axis, the locally resolved flux and the deposition of silica particles on the membrane. Throughout the results section, black and white images present NMR images without any shown information of velocity. Colored images and vector maps express flow-MRI results. Chapters 4.1- 4.3 reflect the MRI results gained from pure-water filtration experiments whereas Chapter 4.4 demonstrate the impact of the silica particle filtration on the flow behavior. For the sake of comprehension, Table 2 summarizes the MRI respective data of the figures including plane thickness, location of the plane ("*Shift from center*") and investigated velocity component ("*Velocity*").

*4.1. Interior structure of the sterile membrane filter*

Two dimensional NMR images uncover the interior structure of the sterile membrane filter in every spatial plane (*xy*, *zy* and *zx*, Figure 5). Therein, white color resembles a high NMR signal coming from water protons whereas black color indicates areas with nearly no signal. Due to this contrast effect, we are able to visualize details within the module that are a prerequisite for the fluid dynamic analysis (e.g. the membrane). In Figure 5a the cross-sectional plane (*xy*-plane) depicts design features, such as the splines in the lid and the separated permeate channels. The high contrast favours the visualization of the membrane that separates the permeate channels and the feed channel. The side view (Figure 5b) demonstrates the huge advantage of MRI by accounting for inhomogeneity. Although the module was under pressure, the wetted membrane bends into the feed channel and forms a fold. Thus, with MRI and flow-MRI non-ideal behavior may be resolved and clearly related to geometric features. Figure 5c shows an NMR image of the permeate channel structure incorporating 18 permeate channels as well as a the permeate collector at the end of the module.

Table 2

Figure guideline containing MRI-specific parameters

| Figure [-] | Plane | Thickness [mm] | Shift from center [mm] | Resolution [$\mu m$] | Velocity [-] |
|---|---|---|---|---|---|
| 5a | xy | 6 | 0 | 97.7 x 66.4 | - |
| 5b | zy | 4 | 4.4 | 195.3 x 62.5 | - |
| 5c | zx | 6 | 2 | 97.7 x 97.7 | - |
| 6a | xy | 6 | 16 | 97.7 x 62.5 | z |
| 6b | xy | 6 | 0 | 97.7 x 62.5 | z |
| 6c | xy | 6 | -16 | 97.7 x 62.5 | z |
| 7 | zy | 4 | 4.6 | 195.3 x 62.5 | z + y |
| 8a | zx | 1 | 4.3 | 195.3 x 97.7 | y |
| 8b-1 | xy | 6 | 2 | 97.7 x 62.5 | z |
| 8b-2 | xy | 6 | -14 | 97.7 x 62.5 | z |
| 9 | xy | 6 | 0 | 62.5 x 195.3 | - |
| 10-left | zy | 6 | -4 | 195.3 x 62.5 | z |
| 10-right | xy | 6 | 0 | 62.5 x 195.3 | z |
| 11 | zx | 2 | 5.4 | 195.3 x 97.7 | - |

*4.2. Influence of module design on filtration performance*

Flow-MRI visualized the velocity in *z*-direction for three different cross-sectional planes along the module axis. Velocities range between 30 and -5 mm/s with positive velocities pointing into the plane. Figure 6a shows the plane right at the inlet. Here, the incoming flow impinges the middle spline creating two dominant flows in the middle compartments of the feed. This split in flow causes maximum velocities of nearly 30 mm/s in the center and, more crucial, low back flowing streams (dark blue coloring) in the rest of the feed. The reason for this flow maldistribution lies in the design of the module' s inlet. In the following, we will observe the consequences of this backflow on the filtration performance. Moving along the module in *z*-direction (Figures 6b and c), the overall mean flow in the feed decreases until we experience slow velocities of around 5 mm/s at the end. Since the membrane module operates dead-end the mean flow in the feed has to decrease due to the steadily increasing filtration. This assumption is based on the fact that the overall velocity continuously rises in the permeate channels although the cross-sectional areas of the permeate channels become larger. Usually, permeation decreases along the membrane axis due to pressure loss or dominating effects like concentration polarization and fouling. In the beginning of the module (Figure 6a), the flow in the center permeate channels (#7 − 12) is the lowest but becomes more and more prominent at the end (Figure 6c). Apart from that, it seems that the velocity dies out in the channels #6 and #13 where high perpendicular velocities occur due to the splines. Based upon the maldistribution in the permeate channels, we can assume that the velocity on top of the membrane encounters higher resistances leading to lower flows.

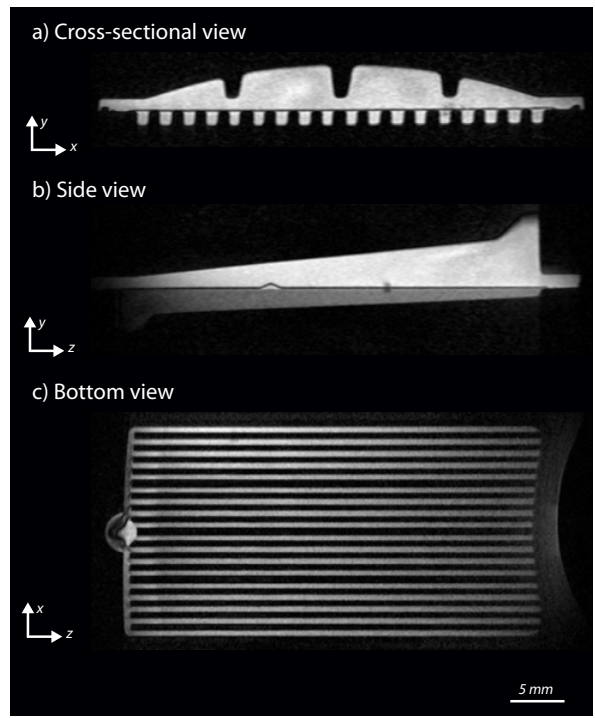

Figure 5: MRI of the membrane filter with its (a) cross sectional view (*xy*-plane), (b) side view (*zy*-plane) and (c) bottom view (*zx*-plane). Images show the high contrast to the membrane. The fold in the membrane (b) reflects the non-ideality influencing the flow in the module.

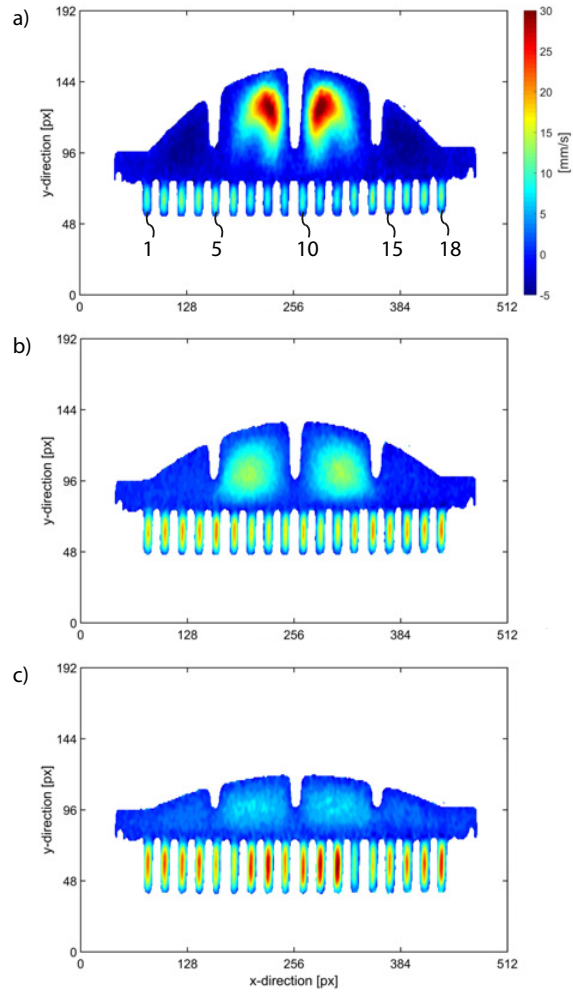

Figure 6: Cross sectional view of the membrane filter at three different axial positions: (a) at the inlet, (b) in the module center and (c) close to the outlet. Flow imaging data show the z-velocity in the feed and in every permeate channel. Along the axial direction the velocity decreases in the feed due to water permeation. In the permeate channels, velocities unevenly increase.

Flow-MRI measurements in the *zy*-plane reveal the flow distribution along the module axis (*z*-direction) as shown in Figure 7. Here, velocities in *y* and *z* direction were superimposed to observe the full velocity field and the fluid's direction towards the membrane. The tail of the mean flow can be recognized by the high velocities ranging from 15 - 30 mm/s and moving from the inlet to the main body of the feed. Interestingly, the incoming flow touches the membrane surface after approximately one third of membrane length. Thus, the first third of the membrane surface remains uninvolved in the filtration process. Figure 7b shows the corresponding vector plot. For better visualization of the flow direction the vector plot is unscaled. At the spot where the flow approaches the membrane surface a backflow emerges and we observe that this part of the membrane shows no velocities in negative $y$-direction, or in other words: at this spot no permeation takes place. This ultimately means that this part of the membrane is inactive and that the active part of the membrane accompanies the mean flow along the $z$-axis. In order to increase the efficiency of the membrane, module optimization would need to change the design of the inlet in the sense that the flow is directed earlier to the membrane surface leading to a higher active membrane area.

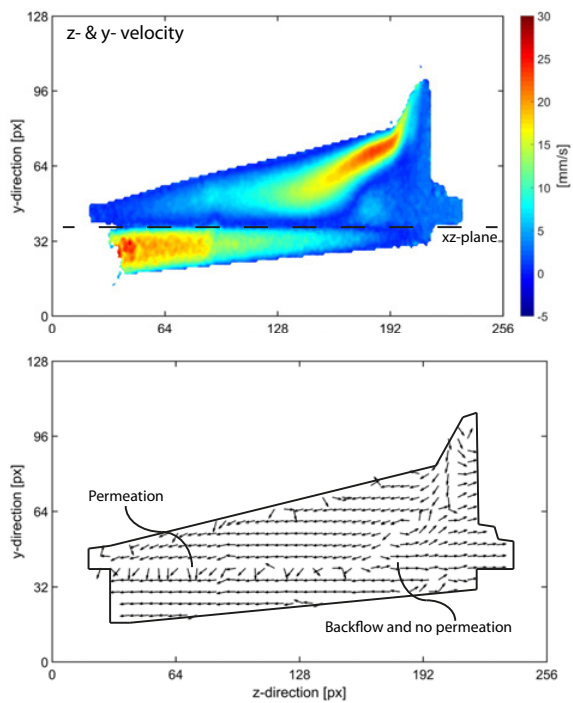

Figure 7: (a) Contour plot and (b) vector plot showing the total velocity ($y$ and $z$ velocity) in the $zy$-plane. Due to the inlet design the inflow is directed towards the membrane at 1/3 of membrane area with highest velocity. In the permeate channels the velocity increases along the module length although the cross-sectional area of the permeate channels increase as well. Vectors are not scaled for better visualization of the flow direction. At the position where the inlet stream touches the membrane a stream forward and a backflow emerge. In the backflow zone no permeation occurs showing the non-activity of the membrane.

*4.3. Locally resolved flux shows the active membrane area*

We now advanced flow-MRI measurements in the *zx*-plane directly below the membrane (see the corresponding dashed line in Figure 7). The contour plot in Figure 8a shows the *y*-velocity that is the volume flow passing the membrane and therefore the flux. Flow-MRI locally resolved the total flux since *y*-velocities were obtained for every single pixel. As a result, we can immediately identify membrane areas with high or low permeation. Regarding the fold in the membrane, the mean flow in the permeate channel largely contributes to the flow in *y*-direction and is by far larger than the flux. Besides that, Figure 8a locally reflects areas with high fluxes but also with nearly zero flux (green coloring). Areas with zero flux are at the beginning of the membrane and in the end (areas marked by dashed lines). In order to validate our local flux measurements, the transported flux was compared to the permeate increase along the module length (*z*-direction). Local flux validation was done for all 18 permeate channels (Table 3). Exemplary, the corresponding *z*-velocities are shown for three permeate channels only (Figure 8b). The *xy*-plane labeled with 1 is the inlet plane and the outflow plane is labeled with 2. The latter has the higher velocities due to the uptake of the permeate flow.

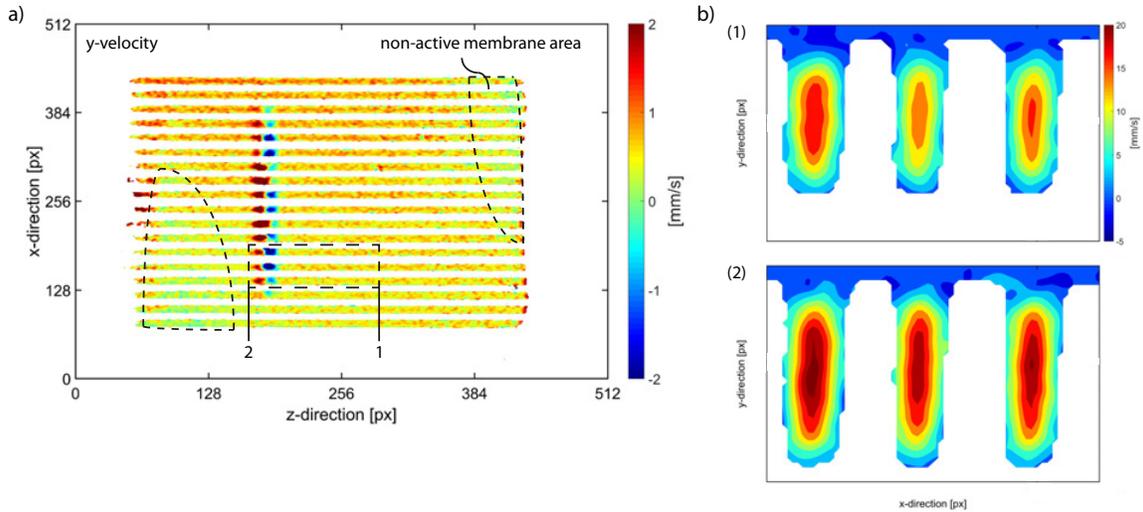

Figure 8: Flow-MRI of the *zx*-plane right below the membrane and the cross sectional areas of marked permeate channels. (left) Contour plot representing the *y*-velocity with highlighted permeate channel area, (right) contour plot of the *z*-velocity for mass balance validation.

A reason for the partially high deviations originate from the fold in the membrane. There, the bulk stream of the flow in the permeate channel has a large impact on the otherwise low flux velocities. On purpose, we chose to include the influence of the fold to account for real applications.

Table 3

Comparison of the measured local flux with the corresponding volume flow between two distinct planes

| Channel [-] | $\Delta \dot{V}$ [$\frac{mm^3}{s}$] | $\dot{V}_{Flux}$ [$\frac{mm^3}{s}$] | Deviation [%] |
|---|---|---|---|
| 1 | 10.0 | 11.7 | 14.1 |
| 2 | 9.5 | 10.7 | 11.5 |
| 3 | 8.9 | 11.7 | 24.3 |
| 4 | 9.8 | 11.3 | 13.4 |
| 5 | 10.6 | 9.3 | 11.5 |
| 6 | 9.8 | 8.8 | 9.6 |
| 7 | 10.9 | 9.8 | 10.1 |
| 8 | 10.0 | 10.4 | 4.0 |
| 9 | 9.3 | 9.4 | 0.9 |
| 10 | 11.4 | 9.7 | 14.6 |
| 11 | 11.7 | 11.4 | 2.1 |
| 12 | 9.5 | 11.1 | 14.1 |
| 13 | 9.0 | 8.9 | 1.0 |
| 14 | 11.1 | 9.7 | 12.6 |
| 15 | 10.9 | 9.7 | 11.4 |
| 16 | 10.0 | 9.3 | 7.5 |
| 17 | 10.2 | 10.7 | 4.2 |
| 18 | 10.9 | 11.3 | 3.1 |

*4.4. Impact of silica particles on cake layer evolution and permeation*

Since silica particles do not induce any MRI signal, these particles are suitable to investigate their influence on the permeation while and after being rejected by the membrane. Figure 9 shows how silica particles deposit on the membrane surface over time. Within the filtration cycle more and more particles deposit across the whole membrane width but show stronger accumulation in the center and below the splines. There, transversal velocities become stronger due to the lower module height. With increasing filtration time the silica deposits become more prominent in the center and then move more and more to the side of the module. This behavior indicates the increasing flow resistance in the center on the transversal flow in $y$-direction. After the filtration ($t = 54 min$), the NMR image shows an uneven deposition of silica particles which, on the one hand, can be explained by the changes in the fluid dynamics and, on the other hand, have been suggested to be a result of gravity driven settling [12].

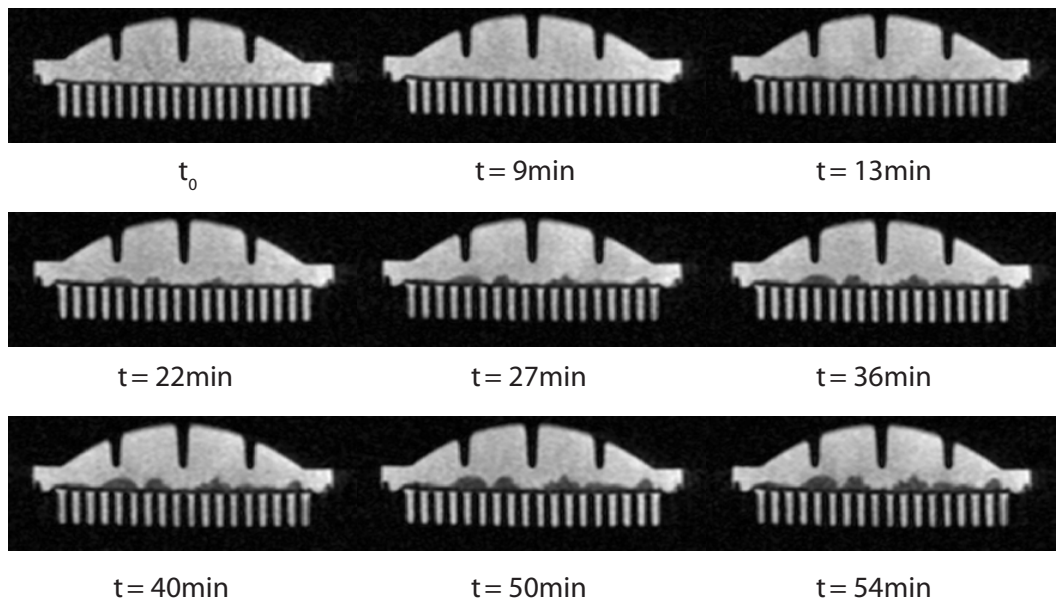

Figure 9: Evolution of cake layer formation on the membrane surface. MRI shows the cross sectional plane (*xy*) in the center of the membrane module. After 13min first silica particle deposits are visible which over time become larger. Largest silica particle deposition seems to develop below the left and right spline.

As expected, the rejected particles compromise the filtration performance of the membrane module. Velocimetry measurements reveal the reduced permeation and the unfavorable changes in feed and permeate flow (Figure 10). Top images present the unaffected flow whereas the bottom images show the fluid dynamics affected by the silica particles. In Figure 10-left the incoming flow is directed to the end of the module avoiding the higher resistances due to the formed cake layer. As a consequence, the overall velocity decreases in the permeate channel. Regarding the cross-section, we can support this statement since the mean flow moves away from the membrane to the lid and to the side of the module, and thus limiting the contact between bulk flow and membrane. As a result, the permeate flow in almost every permeate channel drops to nearly zero except for the two side channels (#1 *and* 18). After one hour of filtration the overall performance of the membrane module heavily suffers from the deposited particles. In order to keep the performance of the membrane module stable for a longer filtration period, means to enhance wall shear rates or mixing at the membrane would be necessary. Especially the fluid dynamics close to the membrane affect the deposition of the particles as shown in Figure 11. Here, darker spots indicate the rejected particles. Except for the side channels (#1 *and* 18) the rest of the membrane becomes darker. A huge amount of silica particles are transported by the backflow close to the inlet forming larger particle deposits. Due to the design of the module' s lid silica particles nicely aligned along two permeate channels to block them.

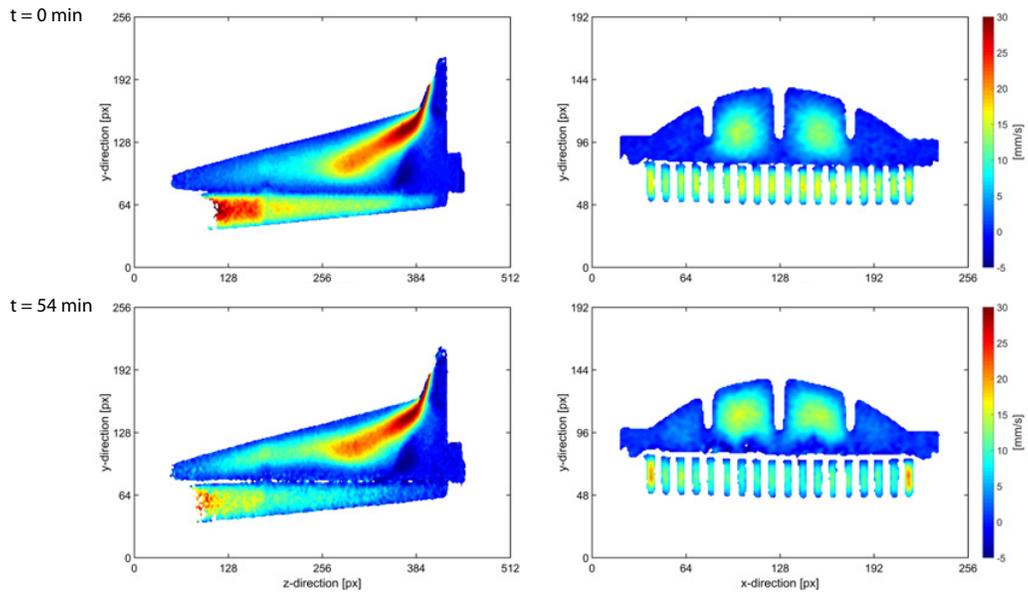

Figure 10: Influence of silica particle deposition on the fluid dynamics. All images reflect the *z*-velocity before and after 54min of filtration. (left) Due to the silica particles the inflow is pushed further to the end of the module and heaved over the accumulated particles (at pixel 256). Additionally, the silica particles deposited on the membrane surface reate a larger contrast which appears in the white slice dividing feed and permeate. (Right) Main flow is forced towards the module lid and to the side of the module. As a result, flow in the inner permeate channels decreases significantly whereas the flow at the sides increases.

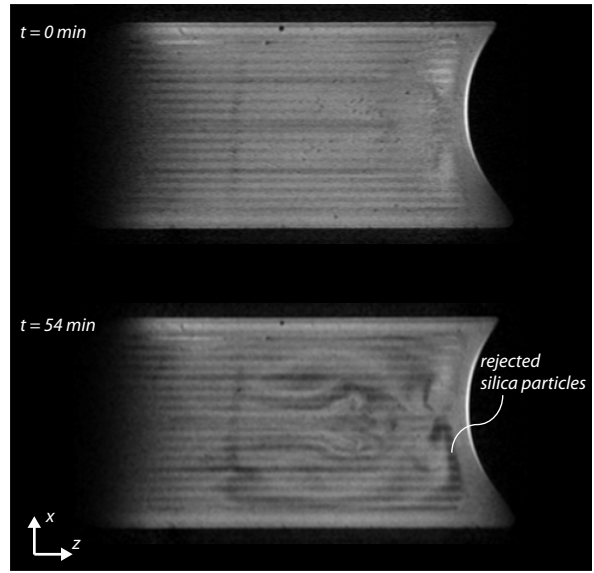

Figure 11: MRI of the membrane surface before and after silica particle filtration. $t = 0$ *min* shows the initially clean membrane surface while $t = 54$ *min* reflects the areas with deposited particles. Darker spots indicate accumulation of silica particles due to the lower spin density.

*4.5. Influence of drag and gravitational force on particle deposition*

The growth of an uneven filtration layer onto the membrane as reported in the previous chapter is unexpected. In fact, one would expect that areas of larger layer thickness would be avoided for permeation due to its higher resistance. As a consequence over time the layer thickness would equalize through the self-regulating permeation behaviour. To evaluate whether or not gravity influences filtration layer build-up as indicated [12], we turned the membrane filter by 180° and repeated the MRI and flow-MRI experiments under the same operational conditions as previously described. Figure 12 presents the time-dependent particle deposition (left) and corresponding velocity evolution (right). At $t_0$, the silica suspension flows towards the membranes bottom-up and the particle deprived water permeates through the initially clean membrane. After approximately 30min of filtration, a cake layer forms on the membrane with valley and hill features at bottom side of the membrane. While filtrating for another 30min thicker cake layers form predominantly at the module outer sides, however not at the locations of thinnest layer and least resistance.

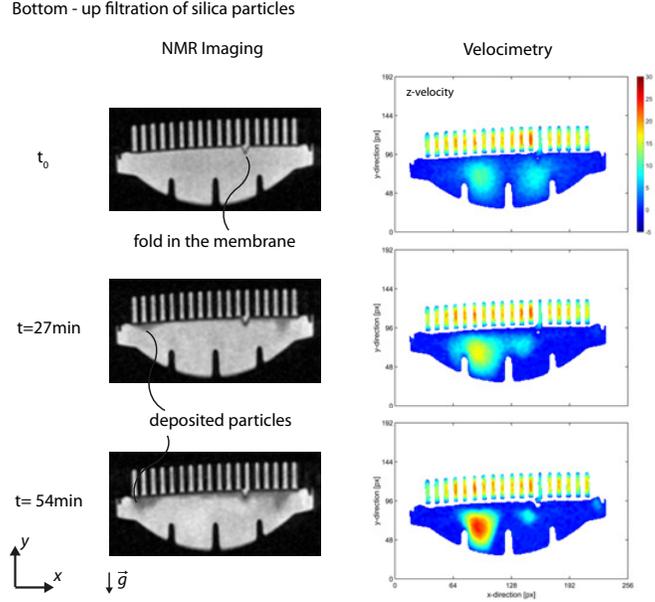

Figure 12: MRI and flow-MRI of bottom-up silica filtration. In this scenario drag forces determine the flow direction of the silica particles towards the membrane.

In combination with the flow-MRI data (Figure 12-right) we can follow the correlation of feed flow and silica particle deposition. At $t_0$ when no particles are rejected by the membrane yet the geometric features of the module determine the feed flow direction that brings the particles to the membrane surface. As soon as a cake layer is formed the feed flow faces an additional resistance and shifts its direction. Here, the feed flow moves to the left part of the membrane with the least cake layer thickness. Not only do the MRI images reflect the deposition of the silica particles on the membrane, they also show that clearly no silica particles deposited at the bottom of the module lid. This ultimately proves our hypothesis that gravitational force does not influence particle deposition on the membrane surface. Using the force balance described above, with $v = 0.175 \frac{mm}{s}$ the forces become $F_D = 4.47 \cdot 10^{-12} N$ and $F_G = 2.43 \cdot 10^{-13} N$, and the resulting Stokes number is $Stk = 0.9 \cdot 10^{-4} \ll 1$. In contrast to our work, Airey et al. [12] showed to have strong particle sedimentation in a hollow fiber membrane during the cross-flow filtration of a 5-wt% silica suspension. However, they used very small Re numbers to enhance the observation of a silica polarization layer and a permeate flux that is two orders of magnitude smaller than ours. Thus, the silica particle experiences only very low velocities in mean flow direction favoring the settlement of the particle due to gravity.

## 5. Conclusion

Using MRI and flow-MRI we investigated the influence of module design on the filtration performance of a commercial combined sterile membrane filter. We found that the fluid dynamics in the feed and permeate of the membrane module are highly heterogeneous. In particular, the locally resolved flux showed active and non-active membrane areas that result from an uneven flow distribution in the feed compartment. Time-resolved imaging of the rejected silica particles reveal how the geometric features of the module affect the deposition of the particles and, in turn, how the deposited particles influence the fluid flow. The results suggest that module design and fluid flow distribution should go hand in hand in order to control time-dependent filtration and rejection performance.

**Acknowledgement**

This work was performed in part at the Center for Chemical Polymer Technology CPT, which is supported by the EU and the federal state of North Rhine-West-phalia (grant no. EFRE 30 00 883 02). Support from the ERC Advanced Investigator Grant ConFluReM (# 694946) has also contributed to the publication. Martin Wiese thanks Ernesto Danieli from Magritek and Stefan Benders from ITMC at RWTH Aachen University for the fruitful discussions. The authors thank Bernd Krause from Gambro Dialysatoren GmbH for the supply of the sterile membrane filters.

## 6. APPENDIX

*6.1. Spin echo pulse sequence*

The NMR Imaging pulse sequence used in our study is Spin-Echo (SE). Figure S1 illustrates the interplay and composition of the pulse sequence based on the used gradient system. As shown in the first line of Figure S1 the SE pulse sequence starts with a slice selected 90° pulse shifting the excited spins out of equilibrium. A subsequent 180° pulse flips the orientation of the dephasing spins and thus compensates for inhomogeneities in the magnetic field. The following lines in Figure S1 represent the excited gradients which are read and phase gradients for spatial encoding, slice selection of the object and velocity gradient for the respective encoding. Here, a pair of gradient pulses ($G_{vel}$) with the same sign are used for velocity encoding since the 180° pulse switches the spins' orientation in-between. Using the duration $\Delta$ of the velocity gradients and the separation time $\delta$, the phase shift is calculated by

$$\phi = \gamma \cdot \delta \cdot G_{vel} \cdot v \qquad (12)$$

for a spin moving with the velocity v [20]. Equation 12 reflects that the encoding method is a direct measurement of the velocity. Therefore, velocity encoding needs a reference image without applied velocity gradient and a second image with activated velocity gradient. Substracting the respective phases pixel by pixel results in the presented quantitative velocity maps.

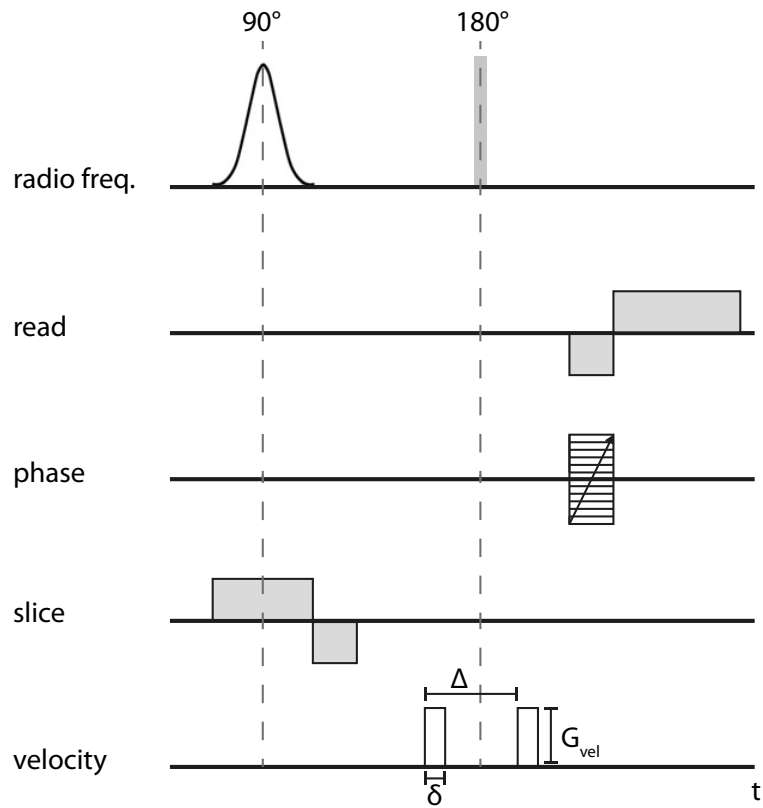

Figure S1: Spin echo pulse sequence

*6.2. Phantom: tube flow for velocity validation*

Preliminary MRI experiments tended to investigate the accuracy of our NMR tomography system for high and low velocity ranges. The need for high and low velocity investigation resulted from the aims of our study within the membrane filter. There, higher velocities are present in the feed flow of the module whereas low velocities were determined for the local flux analysis. Flow-MRI measurements were done in a Plexiglas tube with 7*mm* inner diameter at various volume flows to cover the velocities used in our study. From the cross-sectional flow-MRI images we transferred the data points from the center of the flow to a 1D plot and compared those data with a theoretical, parabolic velocity profile (S2).

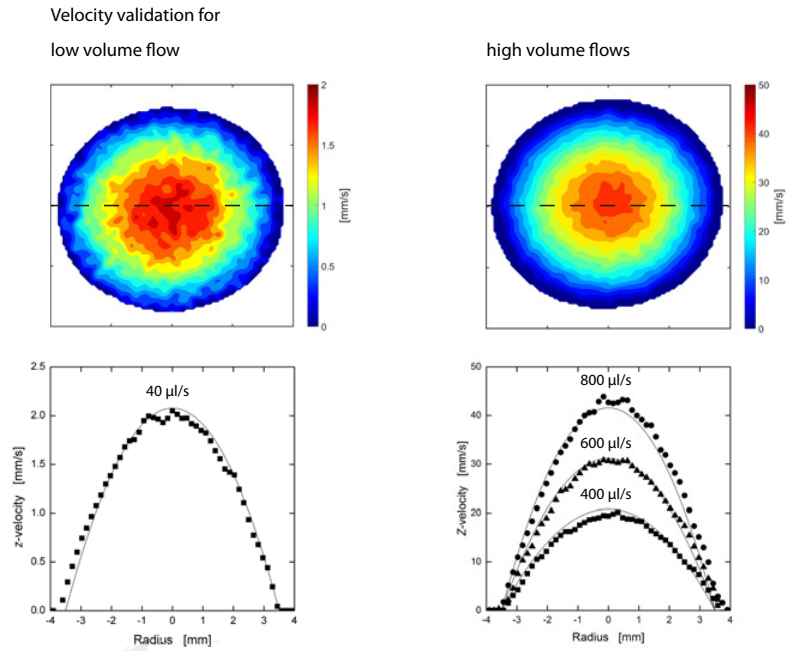

Figure S2: Theoretical and measured velocity profile in tube for MRI validation

MRI data points are illustrated by the black dots whereas the black line represents the theoretical velocity profile in the tube. Theory and experiments are very close with a maximal deviation of smaller than 5%. Especially the MRI data from the low volume flow analysis show that our tomography setup is able to correctly resolve velocities in between 0 and 2 *mm/s* (S2-left).